# A Novel Compact Si-B-N Barrier on Mg-Li Alloys via Plasma Treatment

Y. Chen, J. K. Gao, and X. D. Zhu[*]

*School of Physical Sciences, University of Science and Technology of China, Hefei, Anhui 230026, People's Republic of China*

## ABSTRACT

Mg-Li alloys have attracted much attention due to their superior properties. However, it is a great challenge to improve their inferior oxidation and corrosion resistance. We report a novel Si-B-N ceramic film deposited on the Mg-9.6Li alloy surface as an effective barrier against oxidation and corrosion. The films were deposited by using plasma enhanced chemical vapor deposition from a $N_2$-$B_2H_6$-$SiH_4$ gas mixtures, showing compact structure and adhesive attachment to the alloy surface. The barrier revealed excellent protection against oxidation in humid air for 500 days, and no observable changes were found in immersion test of the 3.5 $wt.\%$ NaCl solution for 10 min. The superior oxidation and corrosion resistance are attributed to the excellent material property of Si-B-N coatings with compact structure via plasma treatment. Moreover, it is found that moderate proportion of $B_2H_6$ in the source gas mixture is beneficial to the protection of alloys, where the hydrogen release reaction nearly disappeared and no bubbles were generated on the surface in the immersion test.

[*]Corresponding author: X. D. Zhu (email: xdzhu@ustc.edu.cn)

# 1. Introduction

Mg-Li alloy is the lightest ($\rho = 1.33 \sim 1.65\ g/cm^3$) engineering alloy system in existence[1,2], which possesses superior properties, such as high specific strength and stiffness, outstanding damping, weak mechanical anisotropy, and excellent electromagnetic shielding performance[3-7]. Although Mg-Li alloys preserve great potential in the fields of aerospace, military, 3C industry, and biomedicine[8,9], there are some serious defects in application, especially the poor corrosion resistance and oxidation resistance in humid air and high-salt environment[10]. The metal impurity elements cause multi-phase electrochemical reactions on the alloy surface which accelerates localized corrosion[10,11]. Unlike stainless steel and aluminum, Mg-Li alloys cannot form self-protection oxide films in the atmospheric environment.

It is desirable to improve surface corrosion resistance through favorable surface treatments of Mg−Li alloys while maintaining the bulk properties of Mg−Li alloys, which has been the subject of intensive research in decades. Many methods have been adopted in surface treatments of Mg−Li alloys, such as anodizing[12], micro-arc oxidation[13,14], electroplating[15], and plasma spraying[16]. However, the protective films formed usually present loose and porous structures caused by localization of electrochemical reactions or by particle formation during film growth procedure. Besides, several chemical conversion coatings are reported to protect the Mg-Li alloys from corrosion[17-19]. Cracks are formed in chemical conversion due to hydrogen release, which provide channels for external corrosion medium. Though these efforts yield valuable progress regarding improving the corrosion resistance of Mg-Li alloys, it remains a great challenge to fabricate high quality barrier layers through desirable surface treatment technology.

Recently, the multi-component ceramics, such as nitrides and carbonitrides of silicon and boron, exhibited attractive corrosion resistance and hardness, which makes them suitable for a new generation of technically important and versatile hard coatings[20-22]. It is of great interests to apply the ternary and quaternary coating on the treatment of Mg-Li alloys to improve the inferior oxidation and corrosion resistance. By contrast with the surface treatments of Mg-Li alloys mentioned above, plasma enhanced chemical vapor deposition (PECVD) by glow discharging possesses unique advantages, such as the low-temperature growth with compact structure due to electronic activation in plasma.

In this letter, we propose a novel Si-B-N ceramic film on the surface of Mg-9.6Li alloys by PECVD. The barrier layers formed presented excellent properties of oxidation resistance and corrosion resistance for the Mg-9.6Li alloys. The barrier revealed excellent protection against oxidation in humid air for 500 days, and no observable changes were found in immersion test of the $3.5\ wt.\%$ NaCl solution for 10 min.

# 2. Experimental

The treatment of Mg-9.6Li alloy was conducted in a device which consisted of a cylindrical chamber with two parallel electrodes inside. The upper electrode was connected to a 13.56 MHz RF power supply, the lower one was grounded as substrate holder, which is 80 mm from the upper one.

The plates of Mg-9.6Li alloys used in experiments were $15 \times 15 \times 5\ mm$ in size. Mg-9.6Li alloy plates were polished by SiC papers to remove the native oxide layers, and cleaned by sonication in acetone and ethanol for 15 min respectively. After pretreatment, the Si-B-N ternary films were deposited on the surface of the Mg-Li alloys in the chamber at a temperature of 473K. The supplied RF power was 200 W. The source gases were mixtures of $N_2$, $B_2H_6$ (2% $B_2H_6$ on $N_2$ dilution), and $SiH_4$ (5% $SiH_4$ on $N_2$ dilution). The chamber gas pressure was maintained at 80 Pa and the growth time was 5 hours in all growth process.

After plasma treatment, we put the uncoated specimens and coated specimens in humid atmosphere for 500 days to evaluate their oxidation resistance. The corrosion resistance was tested in an immersion experiment. The uncoated specimens and coated specimens were parceled by paraffin wax except $10 \times 10\ mm$ area on surface and then immersed in $3.5\ wt.\%$ NaCl solution for 10 min. The surface morphology of the specimens was observed under the XL-30 ESEM scanning electron microscopy (SEM), and the infrared absorption spectrum was obtained by Nicolet 8700 Attenuated Total Reflection Flourier Transformed Infrared Spectroscopy (ATR-FTIR).

## 3. Results and discussion

Fig. 1 shows the photographs of uncoated Mg-9.6Li alloys before/after being put in humid air for 500 days (upper row) and the photographs of coated Mg-9.6Li alloys before/after being put in humid air for 500 days (lower row), where the fraction of $B_2H_6$ to the sum of $B_2H_6$ and $SiH_4$ is 0.67. The alloys appeared metallic luster just after pretreatment, as shown in Fig. 1(a). It can be seen in insert SEM image that the polished alloys already got slightly oxidized with fragmentary oxidation layers, indicating the inferior oxidation resistance of the alloys. After deposition Si-B-N coatings, the specimen revealed faint yellow appearance (Fig. 1(c)). The ceramic films showed compact structure and full coverage as shown in insert image of Fig. 1(c).

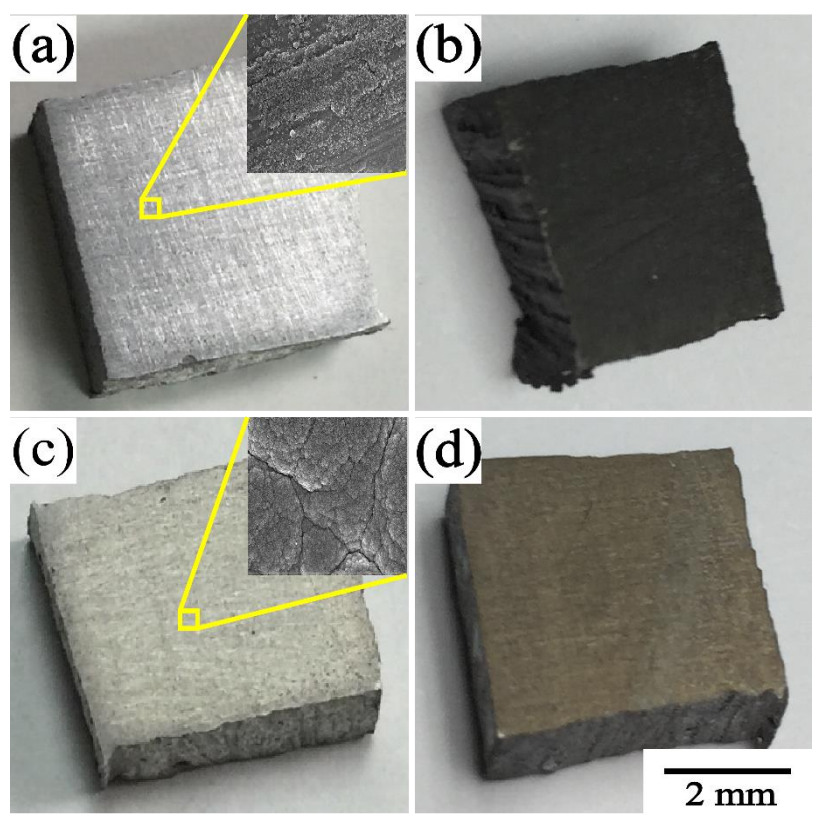

Fig. 1 Photographs of uncoated Mg-9.6Li alloys before/after being put in humid air for 500 days (upper row) and the photographs of coated Mg-9.6Li alloys before/after being put in humid air for 500 days (lower row), the scale bar is $2\ mm$. Insets ($4 \times 4\ \mu m$) show the corresponding SEM images of samples. The fraction of $B_2H_6$ to the sum of $B_2H_6$ and $SiH_4$ is 0.67. (a) Uncoated Mg-9.6Li alloy after pretreatment; (b) Uncoated Mg-9.6Li alloy put in humid air for 500 days; (c) Mg-9.6Li alloy coated with Si-B-N coating; (d) Coated Mg-9.6Li alloy put in humid air for 500 days.

We put uncoated specimens and coated specimens in humid atmosphere for 500 days for the evaluation of sample oxidation. As shown in Fig. 1(b), thick black oxidized layer appeared on the surface of uncoated Mg-Li alloys, which presented poor adhesion and loose structures. This kind of oxidation films is not protective, which is not able to stop further oxidation for Mg-Li alloys.

Different from the uncoated specimens, it is found that Mg-Li alloys coated with Si-B-N coatings exhibited superior oxidation resistance in humid air after 500 days. No observable changes appeared on the surface of the coatings as shown in Fig. 1(d), except that the surface color becomes dimmer compared with Fig. 1(c).

The oxidation experiments demonstrate the treatment of plasmas containing Si, B, N can effectively enhance the oxidation resistance for Mg-Li alloys in humid air. This improvement is attributed to the superior physical and chemical properties of formed Si-B-N films, which are expected to combine the properties of silicon nitride and boron nitride. These films would be resistant to high temperatures and corrosive environments. The multi-component ceramics compounds are commonly deposited using CVD techniques at elevated temperatures. There exists a lot of ions, electrons, and radicals in the plasma. Activated precursor species from the dissociation of source gases, $N_2$, $B_2H_6$ and $SiH_4$, are incorporated into a substrate, leading to film growth with dense and pinhole free structures. The growing process could enjoy a low growth temperature, especially at room temperatures due to electronic activation, which retards the electrochemical reaction on the alloy surface.

We have further carried out the experimental investigations on corrosion-resistance property of uncoated and coated Mg-9.6Li alloys. The specimens are immersed in $3.5\ wt.\%$ NaCl solution for 10 min. The results are shown in Fig. 2.

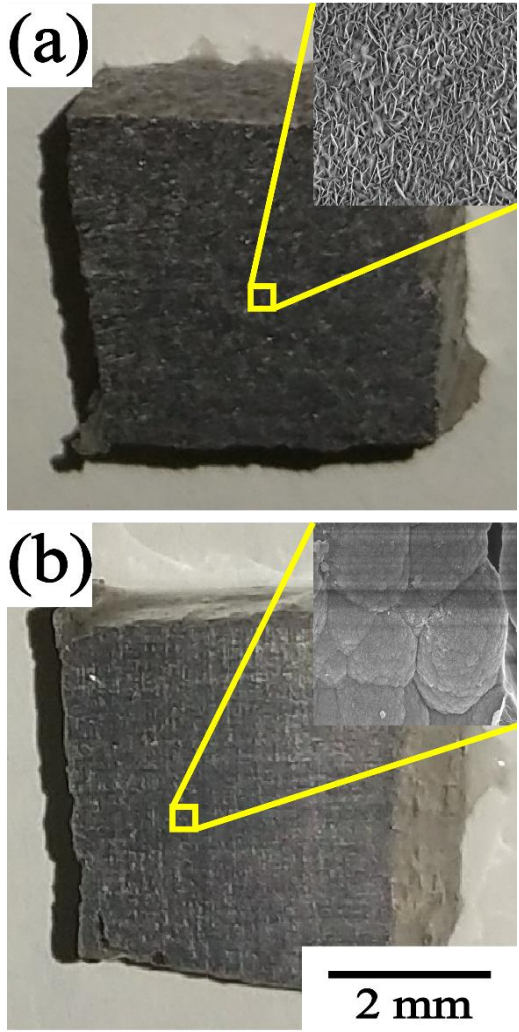

Fig. 2 Uncoated and coated Mg-9.6Li alloys after immersion in $3.5\ wt.\%$ NaCl solution for 10 min and dried in cold air, the scale bar is $2\ mm$. Insets ($4 \times 4\ \mu m$) show the corresponding SEM images of samples. The fraction of $B_2H_6$ to the sum of $B_2H_6$ and $SiH_4$ is 0.67. (a) Photograph and SEM image (insert image) of uncoated Mg-9.6Li alloys after immersion in $3.5\ wt.\%$ NaCl solution for 10 min; (b) Photograph and SEM image (insert image) of uncoated Mg-9.6Li alloys after immersion in $3.5\ wt.\%$ NaCl solution for 10 min.

Fig. 2 shows uncoated and coated Mg-9.6Li alloys after immersion in $3.5\ wt.\%$ NaCl solution for 10 min and dried in cold air. The fraction of $B_2H_6$ to the sum of $B_2H_6$ and $SiH_4$ is 0.67. Bubbles were instantly generated from the uncoated surfaces once immersed by solution. The main composition of the bubbles is hydrogen which is released through electrochemical reaction[23]. After dried in cold air, the sample presented observable corrosion points, as shown in Fig. 2(a). And the SEM image shows the leaflike surface in the insert, suggesting severe corrosion happened on Mg-9.6Li alloy.

In contrast, there were no bubbles generated on the coated alloys in Fig. 2(b), which means the coated alloy owned much better corrosion resistance than the uncoated alloy in salt solution. With a further increase of immersion time, a few small corrosion spots appeared on the coated surface. It is believed that this phenomenon is induced by the inevitable defects formed in growth process of Si-B-N film due to the rough interface feature. One should note the scrubbed marks and textures on the surface of Mg-9.6Li alloys after pretreatment (Fig. 1(a)). The activated precursor species in the plasma tend to assemble around the interface defects, leading to nonuniform growth of Si-B-N films. Therefore, there forms microdefects inside the Si-B-N films during growth process. As immersion time is long enough in high concentration NaCl solution, it is possible for corrosive intermediate (Cl⁻) to penetrate micro defects and reach Mg-Li alloys matrix, causing electrochemical corrosion. At the same time, the hydrogen release cause damages to the Si-B-N films, and forms corrosion spots, which provides convenience for forward corrosion.

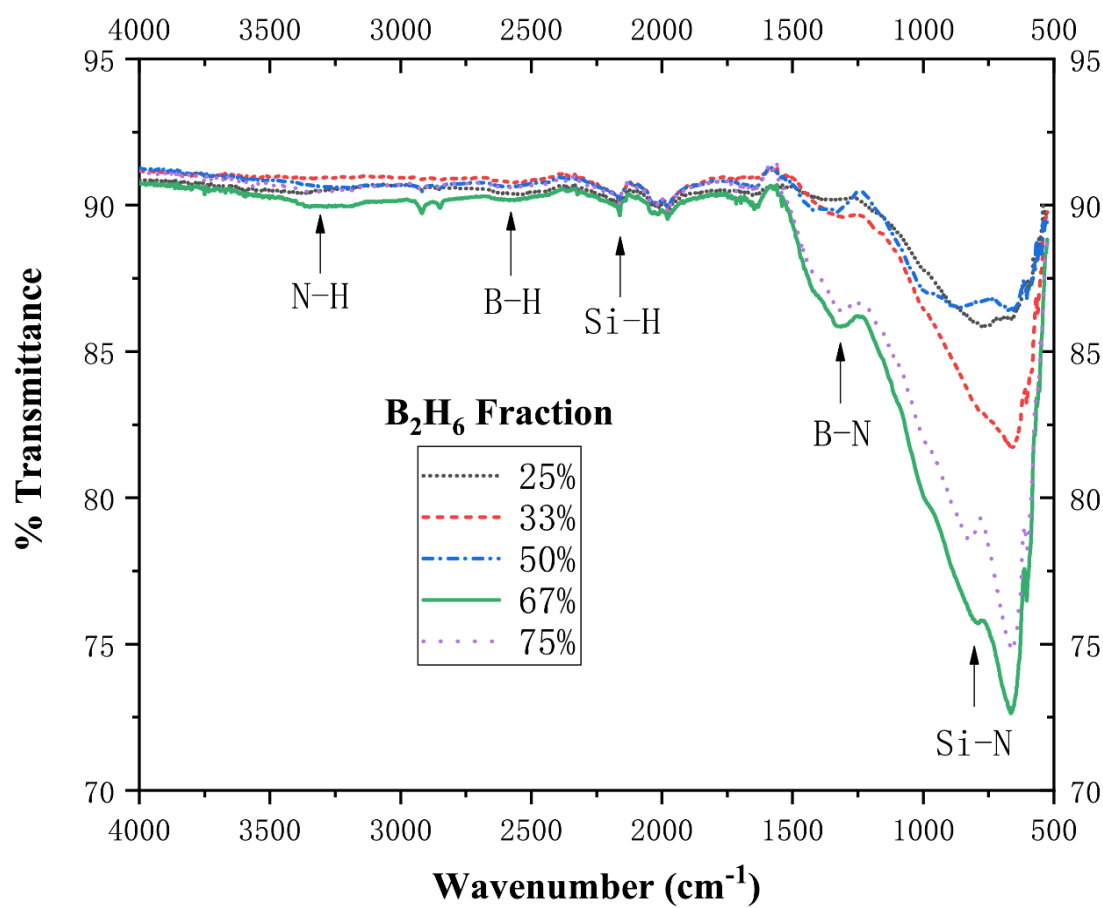

Fig. 3 Infrared spectrums of the Si-B-N films under five typical conditions: the fraction of $B_2H_6$ to the sum of $B_2H_6$ and $SiH_4$ is 0.25, 0.33, 0.50, 0.67, 0.75, respectively.

The fraction of $B_2H_6$ is an important variable affecting gas phase reactions in $SiH_4$-$B_2H_6$-$N_2$ plasmas. The concentration and states of the activated precursor species formed in plasma is changed by adjusting the fraction of $B_2H_6$, which affects further composition and structure of the film. We also experimentally investigated the influence of flow rate of $B_2H_6$ to the sum of $B_2H_6$ and $SiH_4$ on the film structure and corrosion resistance. The flow rate of $B_2H_6$ was set to 0.25, 0.33, 0.50, 0.67, 0.75 respectively while the total flow rate of $B_2H_6$ and $SiH_4$ was kept constant to 18 sccm.

Fig. 3 shows the corresponding infrared spectrums of the Si-B-N films. The strong and broad absorbance around 890 and 1320 cm$^{-1}$ are assigned to be Si-N and B-N lattice vibrations, respectively[22]. The B-N band is similar to that of hexagonal boron nitride, which indicates that Si atoms are substituted by three-coordinate B atoms. The absorbance bands around 2200, 2500, and 3400 cm$^{-1}$ are assigned to Si-H, B-H, and N-H stretching vibrations, respectively.

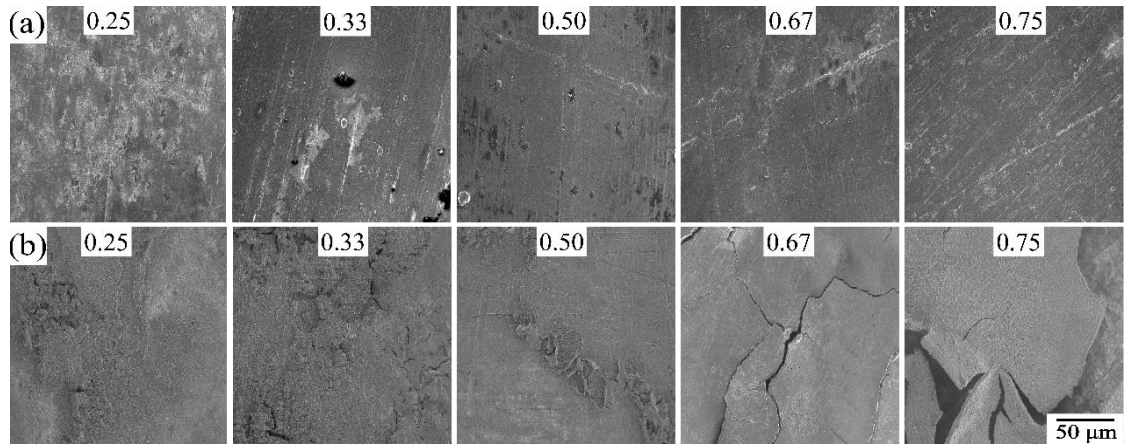

Fig. 4 SEM images of the Si-B-N films before/after immersion in $3.5\ wt.\%$ NaCl solution for 10 min under five typical conditions: the fraction of $B_2H_6$ to the sum of $B_2H_6$ and $SiH_4$ is 0.25, 0.33, 0.5, 0.67, 0.75, respectively. (a) SEM images of the Si-B-N films; (b) SEM images of the Si-B-N films after immersion in $3.5\ wt.\%$ NaCl solution for 10 min.

Fig. 4(a) shows SEM images of the Si-B-N films under various fractions of $B_2H_6$. At low fractions of $B_2H_6$, the films presented discontinuous surfaces. As the fraction of $B_2H_6$ increases, the films tended to become more continuous and complete. Fig. 4(b) displays the influence of the $B_2H_6$ fraction on corrosion resistance in $3.5\ wt.\%$ NaCl solution for 10 min. Compared with the corrosion of uncoated alloys shown in Fig. 2(b), the corrosion resistance of coated alloys with all $B_2H_6$ fractions was improved obviously.

When the fraction of $B_2H_6$ is relatively low, small bubbles were generated slowly. Corrosion spots occurred on the coatings and fragments began to peel off, which was due to the incomplete coverage for discontinuous coatings. When the fraction of $B_2H_6$ becomes too high, small bubbles still existed. Cracks appeared on the surface as shown in Fig. 4(b). It is reasonable to infer that the release of hydrogen induces the film cracks. At moderate proportion of $B_2H_6$, no bubbles were generated on the surface and the hydrogen release reaction nearly disappeared, which is beneficial to the protection of alloys.

## 4. Conclusion

In summary, we propose a novel ceramic (Si-B-N) film to the surface of Mg-9.6Li alloys through plasma enhanced chemical vapor deposition method using SiH$_4$-B$_2$H$_6$-N$_2$ mixture, as a barrier to prevent the Mg-Li alloys from oxidation and corrosion. The films showed compact structure and adhesive attachment to the surface of Mg-9.6Li alloys, effectively improving the oxidation and corrosion resistance of the specimen. The Si-B-N coatings revealed excellent protection against oxidation in humid air for 500 days and no observable changes were found in immersion test of the $3.5\ wt.\%$ NaCl solution for 10 min. The superior properties are attributed to the excellent corrosion resistance of Si-B-N coatings with compact and pinhole free structure via plasma treatment. FTIR experiments shows the obvious Si-N and B-N lattice vibrations, which indicates that Si atoms are substituted by three-coordinate B atoms. It is found that the fraction of $B_2H_6$, is an important variable affecting structure and property of the formed films, and the hydrogen release reaction nearly disappeared and no bubbles were generated on the surface at moderate proportion of $B_2H_6$, which is beneficial to the protection of alloys.

## Acknowledgement

The data that support the findings of this study are available from the corresponding author upon reasonable request. This work was supported by the National MCF Energy R&D Program of China under Grant No. 2018YFE0301102